\documentstyle[aps]{revtex}
\input epsf.sty
\begin{document}
\newcommand{\slq}{\slash\hspace{-2mm}q}
\newcommand{\slp}{\slash\hspace{-2mm}p}
\newcommand{\slt}{\slash\hspace{-2mm}t}
\title{Tensor Analyzing Power and Current Conservation
      in the $\vec{d}+p \rightarrow ^3$He$ + \gamma$ reaction}
\author{S. Nagorny$^{1,2}$, A.E.L. Dieperink$^2$}
\address{$^1$NIKHEF, P.O. Box 41882, 1009 DB Amsterdam, The Netherlands}
\address{$^2$KVI, Zernikelaan 25, NL-9747 AA, Groningen, the Netherlands}
\maketitle
\begin{abstract}
The tensor analyzing power in  $p+ \vec{d}$ radiative capture is
investigated in a covariant and gauge invariant approach.The
results of the present calculation are compared with those of
other theoretical approaches and with experiment. Predictions for
$A_{yy}$ at $E_d = (150-350)$ MeV are given. At forward and
backward angles $T_{20},$ $ T_{22},$ and $A_{yy}$ are sensitive to
the effects of current conservation. The interplay between effects
coming from the nuclear and current structures is discussed.
\end{abstract}
\section{Introduction}
Radiative capture of the tensor polarized deuterons by protons,
and in particular the determination of the tensor analyzing power
(TAP) $A_{yy}$ (or the $T_{20},$ and $T_{22}$ components), has
been proposed [1,2] as a unique source of information on the
tensor component in the $NN$ interaction, which is responsible for
the $D$ waves in  few-body systems. A number of full Faddeev
calculations have been performed during the last several years in
which the three-nucleon dynamics in the bound and the continuum
states is taken into account exactly (see [2-3], and references
there). Since most of these were restricted to intermediate angles
(i.e. near $90^o$ in the c.m.s.) and the low energy region, mostly
only $E1$ transitions were considered [2], assuming  magnetic
dipole and electric quadrupole transitions to be small (as  is the
case for the cross section). The role of high multipole
transitions in $A_{yy}$ was examined only recently [4], although a
first estimate of $M1$, $E2$ transitions and their importance was
mentioned earlier in [5]. In general, these multipoles appeared to
be important even near $90^o$ for $E_d > 30-40$ MeV, and at
forward or backward angles at all energies. These calculations
confirmed the strong sensitivity of the TAP to the $D$ states, and
also showed that there are important effects from initial state
interactions (ISI) in the three-nucleon continuum.
\par
 The reaction mechanism of the radiative capture may be divided
into two main classes: ``external" and ``internal" radiation,
corresponding to the emission of the photons from external  and
internal lines (or vertices), respectively. As  was mentioned, the
ISI and, therefore,  internal radiation in general (of which the
ISI is an essential part), is very important in TAP calculations
[2-4]. Therefore we expect that meson exchange currents (MEC),
reflecting another part of the internal radiation,  also play an
important role (as they do in the cross section).  Moreover, both
types of radiation are not independent, they are connected by the
requirement of gauge invariance [6,7]. And only an exactly
conserved nuclear current guarantees a consistent treatment of the
one- and many-body mechanisms [6] with a correct balance of the
internal and external radiation [7]. In this connection we note
that  previous calculations of TAP did not satisfy current
conservation, although some effects from MEC were partially
included in [2-5] according to Siegert's hypothesis. An additional
iso-vector MEC, which is gauge invariant itself, was included in
[5] explicitly for M1- transition.
\par
In this paper we calculate TAP for the $\vec{d} + p \rightarrow$
$^3$He + $\gamma$ reaction with tensor polarized deuterons, using
the covariant and gauge invariant approach [6] developed earlier
for photo(electro)- disintegration of the few-body systems [7].
Such an approach based on the effective field theory methods
allows one to satisfy the fundamental requirements of covariance
CPT- and gauge invariance on both ``tree" and ``loop" levels, and
at the same time to include intra-nuclear dynamics, final
(initial) state interaction and effects from MEC. It means, that
one- and many-body reaction mechanisms, corresponding to
one-particle reducible (pole type) and irreducible (contact type)
diagrams, are taken into account consistently with nuclear
dynamics and their balance in the total gauge invariant amplitude
is guaranteed by the exact conservation of the total nuclear
electromagnetic (EM) current. This approach was successfully
applied to various observables in the EM interactions of the
few-body systems [7-11]. The unpolarized cross sections in the
$d+p \leftrightarrow ^3$He$ + \gamma$ reaction as well as  the
cross section and structure functions in the $^3$He$(e,e'd)p$
channel, which are the most sensitive to  final state interaction
(FSI) [9], were described well [7,10]. Although, non-relativistic
Faddeev calculations also describe the cross section well, some
deviations in the description of the structure functions at high
transferred momenta, were found [10]. Therefore one expects that
spin-observables are even more sensitive to the effects of
Lorentz-covariance and current conservation, since they involve
the interference of different current components (see, for
instance, [12]). That is why it is of  interest to study the
results for TAP in the framework of the covariant and gauge
invariant theory, including all multipole transitions.
\par
In comparing with various approaches, it is convenient to distinguish:
\par
1. Quantum-mechanical approaches, in particular full Faddeev
calculations, take into account nuclear structure and ISI (or FSI)
exactly, through the solution of the non-relativistic Faddeev
equation for bound and continuum state wave functions, preserving
orthogonality of the initial/final states. However, there are two
main problems: i) gauge invariance is violated, since the mesonic
exchange effects, which are implicitly included in the NN-
interactions, are not included in the EM interactions, and
therefore, there is no balance between one- and many-body EM
current operators in the total amplitude (in the restricted
low-energy region and for not too extreme angles, a special choice
of the EM operators allows one to shift a gauge arbitrariness to
high-multipoles, assuming they are not important; but this
procedure does not reflect a real gauge freedom, since the gauge
is fixed for the main multipoles by some ``good choice" of the
gauge function, and that is evidently not enough since physical
results should not depend upon the choice of the EM operators at
all); ii) the non-relativistic structure of EM operators, which
are defined in the nucleon rest-frame, requires some corrections
in the rest-frame of the nucleus ($^3$He), in which frame  the
Faddeev equation is solved and the calculation of the observables
is performed (an additional Lorentz boost of the operators leads
to a well-known problem: in the $instant$ $form$ of dynamics the
corresponding Poincare generators contain the interaction).
\par
2. The field theoretical approach allows one to combine the
requirements of covariance and gauge invariance with the
accounting for the intra-nuclear dynamics, i.e. ISI/FSI and MEC
effects, even for phenomenological NN interactions [6]. However,
the complicated structure of the relativistic equations for the
n-point Green function (especially in the case of 3- and 4-body
systems) and a large number of various ``loop" (and vertex)
corrections, connected with the inclusion of the regular part of
the hadronic $T-$ matrix (in the case of full ISI/FSI), makes it
practically impossible for applications at present. That is why we
use  a compromise [7] which, on the one hand, preserves all above
mentioned fundamental properties of the theory, and, on the other
hand, allows one to use all achievements of the traditional
nuclear physics, such as solutions of the Faddeev equation, and to
define various 3- and 4-point nuclear vertex functions. Of course,
in this way we loose some relativistic effects (like the small
``negative-energy" form factors in nuclear vertices, which are
unknown in any case, and very often do not contribute to the main
transitions due to the gauge constraints [14-15]), but we preserve
a covariance and current conservation, providing a correct balance
of the different reaction mechanisms in the total gauge invariant
amplitude and, as a result, a correct connection between the
(unobservable) wave function and physical amplitudes.
\par
Details of the field theoretical approach to  EM interactions of
compound systems, including different ``minimal" methods leading
to the same results for the contact currents on the tree and loop
levels, may be found in [6]. All details concerning the $\gamma+
^3$He$ \leftrightarrow p d$ reaction, in particular, are given in
[7]. In this work we only briefly indicate how to derive the gauge
invariant EM amplitudes from the 3- and 4-points Green functions
($tdp$-, $dpn$- and $tppn$- vertices) using ``minimal photon
insertions" into the external/internal hadronics lines and strong
form factors.
\par
\section{Covariant Structure of the Gauge Invariant Amplitudes}
First we introduce the general form of the covariant $^3$He
$\rightarrow p d$ vertex ($t = p+d$):
$$A_{\nu}(p,d,t) = A^0_{\nu} + (\slp-m_p)
A^{-,p}_{\nu} + A^{-,t}_{\nu} (\slt-m_t) +
(\slp-m_p) A^{-,pt}_{\nu} (\slt-m_t).  $$
Only the first term has a quantum mechanical analogue and can be
expressed (in the laboratory system) through the overlap integral
of the $^3$He and $^2$H wave functions. The next three terms
represent so-called ``negative-energy" components which do not
have a non-relativistic analogue. The last term does not
contribute at the tree level, which consists of the pole graphs
with only one off-shell particle. We discard all possible, but
unknown, negative-energy form components from the beginning.
Moreover, the contributions from $A^{-,p}_{\alpha}$ and
$A^{-,t}_{\alpha}$ terms in any case will be cancelled in the main
Dirac (charge) part of the total conserved current in a model
independent way (by the corresponding counter terms coming from
the contact current) due to the gauge constraint for the 4-point
Green function (see [14,15] for details).
\par
The general Lorentz structure of the $A^0_{\alpha}(p,d,t)$
operator, consistent with the CPT- invariance, is (see [7,13])
$${A^0_{\nu}(p,d,t) = \{\gamma_{\nu} D + p_{\nu} B +
d_{\nu} C \} \gamma_5} \eqno(1)$$
\par
However, to eliminate unphysical contributions from the
longitudinal components of the wave function of a (virtual) spin-1
particle in a covariant way, we have to satisfy the additional
condition (in the same way as for the $dpn$ vertex [6,16]):
$${d_{\nu} \ \bar{u}(p) A^0_{\nu}(p,d,t)  u(t) = 0.}\eqno(2)$$
Note, in the case of the gauge fields unphysical (longitudinal)
components (additional polarization states of the vector field) do
not contribute automatically due to coupling to the conserved
currents (see an example in [14]).
\par
Applying condition (2) to eq.(1), we present the $tdp$- vertex in
the following form, including two independent invariant form
factors, $D,B$, which are related with the $S$ and $D$ components
of the overlap integral of the $^3$He and $^2$H wave functions
[7]:
$$G(t,p,d)=\bar{u}(p) A_{\nu}^{0}(t,p,d;-k^2) u(t)
\chi_{\nu}^{\star}(d)$$
$${A^0_{\nu}(p,d,t) = \{ \gamma_{\nu} D(-k^2) + p_{\nu} B(-k^2) +
d_{\nu} [b B(-k^2) + a D(-k^2)]\}\gamma_5.}\eqno(3)$$
Here $u(p)$ is the Dirac bi-spinor, $\chi_{\nu}(d)$ is the
polarization vector (wave function) of a spin-1 particle
(deuteron), while $t,p,d$ are the 4-momenta of the $^3$He, proton
and deuteron, respectively ($t=p+d$), and $k$ is the relative $pd$
momentum. $\gamma_{\mu}$ is a Dirac $4 \times 4$- matrix, and
$\gamma_5 = i \gamma_0 \gamma_1 \gamma_2 \gamma_3$. All
information on the nuclear dynamics is absorbed by the strong
covariant form factors $D(-k^2)$ and $B(-k^2)$, which are
functions of relative momentum $k^2$ (see below). The coefficients
$a$ and $b$ are given by $a=(\sqrt {t^2} +\sqrt {p^2} )/d^2$ and
$2 b= 1-(t^2-p^2)/d^2$.
\par
Note, in ref. [17] a simplified (without physical background)
$tdp$- vertex, expressed in terms of $two$ form factors and the
relative momentum $k$ only, which is inconsistent with the
eqs.(1)-(2), was used. In accordance with CPT- and Lorentz
invariance the correct $tpd$- vertex contains $three$ strong form
factors, as in eq.(1), and to reduce their amount an additional
physical condition should be applied: it cannot be done in an
arbitrary way.
\par
In the second step we need to distinguish $reducible$ ($A^0_{\nu}$)
and $irreducible$ ($A^{irr}_{\nu}$) vertices (see [6,15], for instance):
$${D^{(0)}_{\mu \nu}(d) \ S_0(p) \ A^0_{\nu}(p,d,t) \ S_0(t) \ \ = \ \
D_{\nu \mu}(d) \ S(p) A^{irr}_{\nu}(p,d,t) S(t)},\eqno(4)$$
where $D^0_{\mu \nu}(d)$ and $S^0(p)$ are free Feynman propapagators of
the vector and spinor particles, while $D_{\mu \nu}(d)$ and $S(p)$ are
dressed  propagators including the mass-operator (or self-energy
part). The 3-point reducible/irreducible vertices (on the left/right
hand sides of (4)) satisfy different equations, containing different
intermediate states: either free Feynman or fully renormalized propagators.
Therefore, it is essential that the selected (from the beginning) type of the
vertices (in the left- or right-handside of (4)) should be used to describe
different reaction channels consistently [7,15].
\par
In the next, third, step we derive the 4-point EM Green function
from the strong 3-point Green function ($tdp$ vertex). This
consists of the ``minimal photon insertions" into all external
hadronic lines and vertices in eq.(4). It is evident, that such a
procedure may be performed using either the left- or right-hand
side of eq.(4). In the first (left-hand side) case we obtain a
conserved current in terms of reducible vertices and free Feynman
propagators, while the second (right-hand side) one generates a
conserved current in terms of  irreducible vertices and full
renormalized propagators. Clearly  the physical amplitudes will be
equal only if the corresponding currents are conserved [6,15].
 \par
 To get the simplest structure of the total
current we derive a conserved  current starting from the left-hand
side of the identity (4). This means that we obtain a conserved
current in terms of $reducible$ vertices and $free$ $Feynman$
propagators, while all self-energy parts of the 2-point Green
functions are already included into the definition of the
vertices. Further we omit the $simbols$ $0$ at the free
propagators and reducible vertices.
\par
Following [7], we get a total conserved current on the tree level
which consists of the $external$ and $internal$ radiation parts:
$${J^{tot}_{\mu \nu}(p,d,q)= J^{extern}_{\mu \nu}(p,d,q) +
J^{intern}_{\mu \nu}(p,d,q)}.\eqno(5)$$
The external radiation (fig.1(i)-(iii)) includes the proton,
deuteron, and $^3$He poles (irregular part of the amplitude,
generated by the photon insertions into all external lines):
$${J^{extern}_{\mu \nu} = F_{\mu}(p,p') S_{0}(p') A_{\nu}(p',d,t) +
A_{\lambda}(p,d',t) D^{(0)}_{\lambda \beta}(d') F_{\mu \beta \nu} (d',d)
+A_{\nu}(p,d,t') S_{0}(t') F_{\mu}(t',t)}\eqno(6)$$
The internal radiation mechanism (fig.1(iv)) corresponds to a
contact current (the regular part, which does not possess
pole-type singularities). It is generated by the photon insertion
directly into the strong $tdp$ vertex and phenomenologically
accounts for the photon radiation from the (virtual) ``mesonic
sector" (in the mesonic theory of the strong interaction):
$${J^{intern}_{\mu \nu} = \int^1_0 {d \lambda \over \lambda}
{d \over d q_{\mu}} \{ z_p A_{\nu}(p-\lambda q,d,t'-\lambda q) + z_d
A_{\nu}(p,d-\lambda q,t'-\lambda q)\}}\eqno(7)$$
Here $p' = p-q$, $d'=d-q$, $t'=t+q,$ and $z_{p(d)}$ is the charge
of the proton (deuteron). The relative variables $k_{p(d)}$ and
$k$ (arguments of the strong form factors $D$, $B$) in the
channels with proton (deuteron) and $^3$He poles are defined as
follows (see [7] for details):
$$ k = \eta_d p - \eta_p d \ , \ \ \ k_p = k - \eta_d q \ , \ \ \
k_d=k+\eta_p q $$
$${k_{p \lambda } = k - \lambda \eta_d q \ \ ; \ \ \ \ \ \
k_{d \lambda } = k + \lambda \eta_p q}$$
$$\eta_p=(pt')/t^2 \sim \mu/m_d \ \ \ ; \eta_d=(dt')/t^2 \sim \mu/m_p
\ \ \ ; \ \ \ \mu = m_p m_d/(m_p+m_d)$$
It is important that using these equations the integration in (7)
may be performed $independent$ of the explicit form of the strong
form factors $D(-k^2)$ and $B(-k^2)$. (see, for instance, [6,7]
and [14,15]). Omitting all terms proportional to $q_{\mu}$ and
$d_{\nu}$ (these will not contribute after the contraction with
the polarization vector of the photon and/or deuteron: $q_{\mu}
\epsilon_{\mu}(q)=0$, $d_{\nu} \chi_{\nu}(d)=0$), we convert the
``integral" form (7) of the internal radiation amplitude into the
identical ``differential" form ($R_{\nu}(k) = \{\gamma_{\nu} D(k)
+ p_{\nu} B(k)\} \gamma_5$):
$${J^{intern}_{\mu \nu} = {k_{\mu} \over kq}
[z_p R_{\nu}(k_p) + z_d R_{\nu}(k_d) - (z_p+z_d) R_{\nu}(k)] -
g_{\mu \nu} [a D(k_p) + B(k_p) + b B(k_d)]
\gamma_5}\eqno(8)$$
It was shown a long ago (see, for instance, [6,7] and/or [14,15])
that the transition from the ``integral" form (7) of the contact
current to the ``differential" one (8) cannot depend upon the
explicit form of the strong interaction due to Lorentz invariance.
 Expanding the functions $R_{\nu}(k_{p,d})$ near the $k^2$- point at
$kq \rightarrow 0$, one can check an important property of the
internal radiation amplitude: it has no pole-type singularities
[7].
\par
The 3-point EM vertices satisfying the Ward-Takahashi identities (WTI) are
(see [6,12,14]):
$${F_{\mu}(p,p') = F_1(q^2) \gamma_{\mu} + {F_1(0) - F_1(q^2) \over q^2}
\slq q_{\mu}
- {\sigma_{\mu \nu} q_{\nu} \over 2 m} F_2(q^2),}\eqno(9)$$
for the interaction with a spinor field, and
$$F_{\mu \alpha \beta}(d,d') = -(d+d')_{\mu}[g_1g_{\alpha \beta}-{g_3
\over 2m_d^2}
(q_{\alpha} q_{\beta} -{q^2\over 2} g_{\alpha \beta})]  +
(z_d-g_1) {d'^2-d^2 \over q^2} q_{\mu} g_{\alpha \beta} $$
$$ -{2 g_2 (g_{\mu \alpha} q_{\beta} - g_{\mu \beta} q_{\alpha}) +
g_3 {d'^2-d^2 \over 4 m_d^2} (g_{\mu \alpha} q_{\beta} +
g_{\mu \beta} q_{\alpha} - g_{\alpha \beta} q_{\mu}),}\eqno(10)$$
for the interaction with a vector field.
The EM form factors are normalized by the conditions:
$$F^{p(t)}_1(0)=z_{p(t)} \ ; \
F^{p(t)}_2(0)=\kappa_{p(t)}-z_{p(t)} \ ; \ g_1(0)= z_d \ ; \
g_2(0)=\kappa_d \ ; \ g_3(0) = 2g_2(0)-g_1(0)+Q_d,$$
where $z_{p(t)}$ and $\kappa_{p(t)}$ are the charge and magnetic
moment of the proton ($^3$He), while $\kappa_d$ and $Q_d$ are the
magnetic and quadrupole moments of the deuteron.
\par
It is easy to check that the vertices (9), (10) satisfy the WTI
also in the half-off-shell case (corresponding wave functions are
implied here)
$${(l-l')_{\mu} F^{p(t)}_{\mu}(l,l')/z_{p(t)} =  S^{-1}(l)-S^{-1}(l') \ ;
\ \ \ (l-l')_{\mu} F_{\mu \alpha \beta}(l,l')/z_d = D^{-1}_{\alpha \beta}(l)
- D^{-1}_{\alpha \beta}(l')}\eqno(11)$$
 Note, that the WTI for the on-shell and half-on-shell $\gamma dd$-
vertices are identical (the same as for the spinor-particles [15])
when the unphysical components of the (virtual) spin-1 particle
(deuteron propagator) are eliminated (by the subsidiary condition
(2)). Indeed, any terms at the right hand side of the ``operator
form" WTI (second eq.(11)) proportional to the (virtual) deuteron
momenta do not contribute, being contracted with the (only)
physical states of the (virtual) deuteron wave function (exactly
in the same way as for the real deuteron). Analogously, for the
(half-off-shell) $\gamma dd$- vertex: any terms proportional to
the (virtual) deuteron momenta do not contribute to the physical
amplitude being coupled to the $physical$ states of the (virtual)
deuteron wave function (deuteron propagator), exactly in the same
way as for the real deuteron.
\par
 Using eqs.(11), one can see that the total current
$J^{tot}_{\mu \nu}(p,d,q)$ from (5) is exactly conserved for any
vertex function ($z_t = z_p + z_d$):
$${\bar{u}(p) q_{\mu} (J^{extern}_{\mu \nu} + J^{intern}_{\mu \nu}) u(t)
\chi_{\nu}^{*}(d) = (z_t -z_p -z_d) \bar{u}(p) A_{\nu}(t',p,d) u(t)
\chi_{\nu}^{*}(d) = 0}.\eqno(12)$$
The current (5)-(7) is the minimal necessary set of the reaction
mechanisms which obeys current conservation. The three terms in
(6) represent the pole-type diagrams, while (7) is the regular
part of the amplitude (or contact current). Only the first diagram
in fig.1(i) (with a proton pole) corresponds to the plane-wave
impulse approximation (PWIA). The second diagram in fig.1(ii)
(with a deuteron pole) corresponds to the crossing-channel (recoil
mechanism), while the third diagram in fig.1(iii) with the $^3$He
pole is part of the ISI/FSI [7,12] (it is the dominant one, at
least near threshold), caused by the pole-part of the hadronic
$T$- matrix of the elastic $pd \rightarrow pd$ scattering in the
initial (or final) state [6]. As it may be seen from  the explicit
form of the contact current (8), its value is determined by the
deviation of the nuclear wave functions (and overlap integrals)
from their long-range asymptotics [6,14]. A contact part of the
amplitude (see fig.1(iv)) corresponds to a many-body current and
effectively (implicitly) accounts for the effects associated with
MEC (in the framework of the meson theory of NN- interaction) [6].
\par
 Although the contact
current does not contain pole-type singularities, it must be
included on the ``tree"-level, since it is the contact diagram
that provides conservation of the nuclear current for the class of
pole graphs. Note, that the requirement of a ``minimal trajectory"
(the path of integration in (7)) fixes the contact amplitude in a
unique way [6,14,15]. Let us specify what we mean, since sometimes
in the literature the ``arbitrariness" of the
procedure/prescription to reconstruct gauge invariance has been
discussed. As it is well known, the requirement of gauge
invariance itself does not allow one to reconstruct a
``transverse" (gauge invariant by itself) part of the contact
current. But it does not mean that an arbitrary gauge invariant
construction can be added to the current (7): any contributions to
the physical amplitude must be based on the concrete physical
mechanisms only. In a phenomenological approach (when we do not
specify the mechanisms of the strong NN- interaction) an
additional (fundamental) condition should be used together with
the gauge constraint to fix the contact current in a unique way.
In a``minimal" field-theoretical scheme [6,14] an additional
requirement of the``minimal trajectory" is implied so that the
``transverse" part may only be equal to zero. Indeed, any non-zero
``transverse" contribution may be directly converted into a
deviation from the ``minimal trajectory" and violation of the
``minimal scheme". In this way a purely ``transverse" part of the
amplitude (which is equal to zero on the tree-level) should be
considered as a correction and must be calculated through the
next-order class diagrams (one-loop corrections) using the same
``minimal scheme" on the level of two-body irreducible graphs (see
ref. [6]). Such a ``step-by-step" approach allows for a consistent
gauge invariant description of the effects associated with FSI and
MEC, selecting diagrams in accordance with one-, two-, ..., n-
particle reducible/irreducible contributions and applying the
``minimal scheme" on each level separately. This is a convenient
approach in case when we do not know the lagrangian, but we do
know all hadronic Green functions (n-point vertices).
 For example: 1) if we know only 3-point nuclear
vertex ($tpd$- vertex in our case) we can generate only tree-level
gauge invariant amplitude; 2) if we know 3- and 4-point nuclear
vertices (i.e. $tpd$-, $tpnn$- and $pdpd$- vertices) we can
generate not only a gauge invariant tree-level amplitude, but also
a new gauge invariant set of one-loop contributions (i.e. that
purely ``transverse part" of the contact current which was equal
to zero on the tree-level); 3) if we know the 3-, 4- and 5-point
vertices (i.e. $tpd$-, $tpnn$-, $pdpd$- and $pdpnn$- vertices) we
can generate an additional gauge invariant set of two-loops
contributions (that is a correction to the pure ``transverse part"
of the contact current associated with two-body irreducible
contributions). This procedure in principle may be continued
selecting three-, four-loop corrections and so on (see [6,7]).
 For our particular reaction a pure
transverse part of the contact amplitude is a correction,
accounting for the regular part of the $pd$- scattering $T$-
matrix, and must be generated through the same ``minimal photon
insertions" into all internal lines and vertices of the
loop-diagrams in the equations for the nuclear vertices. However,
in the present paper we will neglect such a contribution.
\par
In addition to the tree level diagrams we also include the nearest
non-pole contributions (see fig.1(v),(vi)): one-loop corrections
to the current (5) whose singularities are the closest to the
``physical region" after the pole-type diagrams. Diagrams in
fig.1(v),(vi) correspond to the photon production due to the
transformation of the deuteron into a spin-singlet $pn$- pair: $d
\leftrightarrow \gamma + \{pn\}_{S} $ [7]. These cannot be
included at the tree level, since there is no bound state in a
spin-singlet $pn$- system. Such a spin-isospin-flip transition
represents an $isovector$ current, contrary to the $isoscalar$ one
in eq.(5) (see also [8]). Although it also belongs to the
$internal$ $radiation$ class, it evidently cannot be taken into
account by the diagram in fig.1(iv). The isovector current in
fig.1(v),(vi) may be presented in the form [7] ($p_2 = p_1 + q$, \
$t = p_1 + n + p$, \ $n+p_2 =d$):
$${J^{\{pn\}_S}_{\mu \nu} = -i \int {d^4 n \over (2\pi)^4}
\{ Tr [S(p_2) F_{\mu}(p_2,p_1) S(p_1) A_s(p,p_1,n) S^T(n)
\bar{\Gamma}^{dpn}_{\nu}(n,p_2)] \ + \  (p \leftrightarrow n) \}}\eqno(13)$$
The 4-point nuclear vertex $A_s(p,p_1,n)$ of the virtual $^3He
\rightarrow p +\{pn\}_s$ break-up with a spin-singlet $pn$- pair contains
a strong form factor $G_s$ which is a combination of the $S$- and $S'$-
components of the $^3$He wave function [7,9]:
$${\bar{u}(p) \bar{u}(p_1) \bar{u}(n) \ A_s(p,p_1,n) \ u(t) =
G_s \ [\bar{u}(p) u(t)] \ [\bar{u}(p_1) \gamma_5 C \bar{u}^T(n)],}\eqno(14)$$
where $C$ is the charge conjugation matrix.
\par
Using (14) and the explicit form of the $dpn$ vertex $\Gamma^{dpn}_{\nu}$
[16], we get from (13):
$${J^{\{pn\}_S}_{\mu \nu}(p,d,q) = -i \int {d^4 n \over (2 \pi)^4} {G_s
\over n^2-m^2+i\epsilon}{I_{\mu \nu} \over p_2^2-m^2+i\epsilon}}
\eqno(15)$$
$$(4im)^{-1} I_{\mu \nu} = (\mu_p-\mu_n) \Gamma_1 \epsilon(d q \mu \nu)+
(\kappa_p-\kappa_n)[({dn \over m^2} -1) \epsilon(d q \mu \nu) -
2 \epsilon(n q \mu \nu)] \Gamma_1$$
$${+[(\mu_p-\mu_n) \Gamma_2 + (\kappa_p-\kappa_n)(\Gamma_1-\Gamma_4)]
{n_{\nu} \over m^2} \epsilon(d q \mu n)},\eqno(16)$$
where $\Gamma_{1,2...,4}$ are strong form factors in the $dpn$
vertex (without the ``negative-energy" form factors, $\Gamma_3 =
0$). Their expressions in terms of the deuteron wave functions may
be found, for instance, in [6,16]); we also introduced \
$\epsilon(a b c d)=\epsilon_{\mu \nu \alpha \beta} a_{\mu} b_{\nu}
c_{\alpha} d_{\beta}$.
\par
Taking into account that $q_{\mu} I_{\mu \nu} = 0$ independent of
the strong dynamics (due to the fact that all fully antisymmetric
tensors $\epsilon_{\mu \nu \alpha \beta}$ in (16) contain already
the vector $q$), we see that the current (14) is always conserved:
$${q_{\mu} J^{\{pn\}_S}_{\mu \nu}(p,d,q) = 0}.$$
We note that the isovector internal radiation amplitude (13),
$J_{\mu \nu}^{\{pn\}_S}$, is sometimes approximated in the
literature by the introduction of the external radiation diagram
with a ``scalar deuteron" ($d^{\star}$) pole (see ref. [17], as an
example). However, such a simplification is inconsistent, since it
requires additional (free) parameters and/or special assumptions
for the $tpd^{\star}$ vertex (replacing the unbound $d^{\star}$
wave function by a bound one, for instance). As  was found
recently (see [7-9]), the isovector transition (with
spin-isospin-flip) is very important for the cross sections [8,10]
and spin observables [9]. That is why a consistent treatment of
the spin-singlet pairs is essential.
\par
As a last step, to get a numerical result, we need to define all
strong form factors appeared in eqs. (3), (15)-(16). Since
solutions of the corresponding relativistic equations are not
available at the moment, we express all form factors through the
corresponding wave functions (and/or overlap integrals) in the
laboratory system (i.e. in the system where the solution of the
Faddeev equation is obtained), in the same way as it was done for
the form factors in the $d \rightarrow pn$ vertex [16]:
$${({p^2\over 2\mu}+\epsilon_2)^{-1} D(p) \sim I_0
(p)+I_2(p)/\sqrt 2 \ , \ \ \ \ \  ({p^2\over 2
\mu}+\epsilon_2)^{-1} m_p B(p) \sim I_0 (p) + \sqrt 2 {4 m^2 \over
p^2} I_2(p)},\eqno(17)$$
$${({\vec{l}^2 \over 2 \mu} + {\vec{k}^2 \over m} +\epsilon_3)^{-1} \ G_s(l,k)
\ \ \sim \ \ \Psi^S(l,k)+ \Psi''(l,k)},\eqno(18)$$
where $I_{0(2)}(p)$ is the $S(D)$- part of the overlap integrals
between $^3$He and $^2$H wave functions, while $\Psi^S$ and
$\Psi''$ are the fully symmetric and mixed symmetry components,
respectively; $\epsilon_{2(3)}$ is a binding energy in the
two-(three-)body break up channel.
\par
In the case of the isovector transitions (transformation of a
spin-singlet $pn$- pair into a deuteron) the last two diagrams in
fig.1(vii),(viii) do not contribute if we neglect
``negative-energy" form factors ($P$- waves) in the $dpn$ vertex.
We will also neglect ``non-pole" type contributions from the
orthogonal states of unbound $pn$ pairs in a spin-triplet states
which appeared to be too small in a wide kinematical region
[7,8,10], since the main contribution from the spin-triplet $pn$
pair is already taken into account by the deuteron-pole diagram
(see fig.1(ii)).
\section{Numerical Results and Conclusions}
For numerical calculations we use the solutions of the Faddeev
equations with Reid Soft Core (RSC) [18], and Argonne-V18 +
Urbana-IX [19] (V18) interactions. By comparing these we obtain an
estimate of the sensitivity of the TAP to various models of the
nucleon-nucleon potential. Moreover, since the TAP is mostly
determined by the small admixture of $D$ states components (which
is at the level of $8 \%$) in the vertex functions (see [1-4]),
the comparison of the results for different models of NN
interaction with the experiments could provide a separation of the
effects coming from the nuclear dynamics and the structure of the
EM current.
\par
 Fig.2 shows a comparison of the existing calculations for
$T_{20}$ with the experimental results at $E_d = 19.6$ MeV. As one
sees the present results (including all multipole transitions) for
the RSC and V18 interactions are very close; they describe the
experimental points well, reproducing the asymmetric behaviour
around $90^o$ and even the (same) trend at forward/backward
angles.
\par
 The results of ref. [2]
(symmetric around $90^o$), based on the full Faddeev calculations
 for the bound and
continuum state wave functions with the Paris potential, contain
$E1$ transitions only. These describe the experimental points for
$T_{20}$ at intermediate angles only, indicating the need to
include, at least, $M1$ and $E2$ multipoles even at such a low
energy. Comparing the present results and those from ref. [2], one
can see that at $forward$/$backward$ angles in both cases $T_{20}$
 has the same sign:  $positive$ in our case, and $negative$
in the case of ref.[2]. Of course, this is not surprising, since
magnetic multipoles (which are not included in ref. [2]) are
important at extreme (small/large) angles.
\par
 However, a further
comparison with the results reported in ref. [4], which are also
based on full Faddeev wave functions with the Bonn potential (for
the initial/final states), but include all multipole transitions,
contrary to the ref. [2], shows a $positive$/$negative$
``asymmetry" of the $T_{20}$ at $forward$/$backward$ angles.
However, our present gauge invariant calculations (solid- and
dashed-curves in fig.2), which also include all multipoles, do not
show the change of the $T_{20}$- sign with
 increasing the angle from 0$^o$ to 180$^o$. We note that an
 ``asymmetric" behaviour of $T_{20}$
observed in ref.[4] for forward/backward directions cannot be the
result of the inclusion of the full ISI. Our present PWIAS (not
gauge invariant) calculation, i.e. accounting for the proton +
deuteron poles (fig.1i+1ii) in the Coulomb gauge and the
iso-vector current (13) (fig.1v+1vi), quantitatively reproduces
the same ``asymmetric" behaviour of $T_{20}$ with the negative
(positive) signs at forward (backward) angles (we do not show this
not gauge invariant calculation, since it has no physical
meaning). However, inclusion of all other diagrams, which restore
current conservation, makes $T_{20}$ again ``symmetric", as it is
in fig.2, for any $NN$- interactions. Taking into account that
this effect does not depend on the type of $NN$ interaction
(compare our present results for RSC and/or V18 with the ones in
ref.[4]), we suggest that the different behaviour of $T_{20}$ at
extreme angles is connected with the effects of current
conservation which provides a consistent treatment (or a correct
balance) of the $external$ and $internal$ radiations. In other
words: the observed effect appears to be connected with the
structure of the total current, but not with the nuclear
structure, since it exists for various realistic NN models.
\par
 From the comparison of the theoretical curves in fig.2 with the
 experimental
results in the intermediate angle region, where TAP was proposed
to be used as a test of the tensor forces [1,2], we can conclude
that, in general, all realistic potentials (V18, Bonn, Paris and
RSC) contain a reasonable $D$ components at small momenta (which
are tested by the deuterons with $E_d = 19.6$ MeV). Although, the
V18 and Bonn potentials give better agreement with experiment near
$\theta = 90^o$ (where high multipoles, and an exact balance of
the internal/external radiations are not very important at this
particular energy), the accuracy of the measurements is clearly
not enough to make a choice.
\par
In fig.3-5 the present  calculations are compared with the
experimental results at $E_d$ = 29.0; 45.0; 95.0 MeV, as well as
with previous full Faddeev quantum mechanical calculations [4],
where the $Siegert$ $hypothesis$ was used. All figures clearly
indicate an increase of the difference between the covariant and
non-relativistic calculations at forward/backward angles with the
increasing of the
 deuteron energy. Moreover, at high energies (see fig.5) the
 disagreement appears at all angles. Unfortunately, the
existing experimental results cover only intermediate angle
region. Therefore, we would like to stress the need for new
measurements of TAP in wide angle region for any possible
energies. This could allow one to check the sensitivity of TAP to
relativistic effects and current conservation.
\par
 A comparison of the present and previous [4] calculations in the
intermediate angle region shows that both results are very close
at $70^o < \theta < 130^o$, contrary to the forward/backward angle
region (see fig.3,4). This clearly confirms a well known fact [20]
that at low energies ($E_d < 50-60$ MeV) and for the angles near
$90^o$ it is not so important (for any observable, including TAP)
whether external and internal radiations (pole and contact
currents) are included consistently. The matter is: in the above
mentioned conditions the Siegert hypothesis ``works" well enough
(the $E1$- transition dominates) and a ``current conservation" may
be arranged artificially using the continuity equation to express
the longitudinal current components in terms of the charge-
density operators. This corresponds to a ``good choice" of the EM-
operators when the gauge is fixed for the main multipole
transitions, while transfering the gauge-arbitrariness to the high
multipoles [20]. Since the later are not important at small
energies, this makes a non-gauge invariant theory practically
independent of the gauge arbitrariness in these very restricted
kinematics conditions (certainly, such a procedure does not
reflect full $gauge$ $freedom$ for all multipoles). Indeed, as one
can see from fig.5, this is not the case already at $E_d = 95$
$MeV$, where the Siegert hypothesis does not work. The same
situation appears at forward/backward angles, but in this case at
all energies.
\par
 Therefore we may conclude: the comparison of both calculations
(the present one and those from [4]) with the experimental results
at $E_d = 29.0$ and $45.0$ MeV for the middle-angle region near
$\theta = 90^o$ in fig.3,4 shows that in the Bonn and Argonne-V18
potentials the tensor forces are presented reasonably well, in
general. However, it seems that at $29.0$ MeV the $D$ components
are only a bit overestimated (see fig.3) by both potentials,
although, evident lack of experimental points even in the
middle-angle region does not allow to make more complete
conclusions. As for the $E_d = 95$ MeV, even in the intermediate
angle region, TAP strongly depends upon both reaction mechanisms
and nuclear structure (see fig.5), and it is not possible to make
any concrete conclusions about $D$- states when the $external$ and
$internal$ radiations are not included consistently with nuclear
dynamics, in a gauge invariant way. Further measurements of TAP at
high energies could provide an excellent test of the total
conserved nuclear current, and could be used to get a consistent
information about tensor-forces.
\par
 In fig.6 we present a set of predictions for TAP at various
energies: from $E_d = 150$ MeV up to 350 MeV. Such measurements
could be performed, for instance, at KVI, IUCF, TRIUMF, and could
be used as a filter of different theoretical approaches and models
for NN interactions. To show that the present approach predicts,
at least, the absolute values of the cross sections in the above
mentioned energy interval, we compared in fig.7,8 our results
(solid curves) for $^3He +\gamma \rightarrow p + d$ reaction with
the most complete at the moment (for such energies)
non-relativistic calculations [21]. In the diagramatic approach of
J.M. Laget [21] the two- and tree-body MEC (dashed and
dashed-dotted curves, respectively) are included in addition to
the FSI and PWIA diagrams. From a comparison in fig.7,8 with the
experiments at $E_{\gamma}=240$ MeV, and 340 MeV (see references
in [21]) we conclude that the present calculations (solid curves)
describe the existing experimental results on the cross sections
in general. Although, estimation of the loop-corrections,
associated with the regular part of the $pd \rightarrow pd$ $T$-
matrix, would be interesting from the theoretical point, the
experimental uncertainties are still too large and new
measurements are needed.
\par
 To summarize, in the $\vec{d}-p$ radiative capture the TAP
is very sensitive to the effects of current conservation and
consistent treatment of the external and internal radiation
mechanisms at extreme $forward$/$backward$ angles even at small
initial energies. Although, at intermediate angles and at $E_d <
50-60$ MeV, TAP indeed may discriminate between various NN
interactions in accordance with their tensor forces, the exact
conservation of the EM current becomes important in TAP
calculations at all angles starting from $E_d > 70-80$ MeV.
\par
After completion of this work two calculations of the radiative
$pd$- capture appeared [24,25].
\par
In [24] authors used basically the same approach, as was developed
in ref.[7], but with some approximations (see [26] for details).
For instance, a simplified tdp-vertex (as in [17]), depending on
the relative $pd$- momentum only, was used (see eq.(11) of ref
[24]), and the spin-singlet $pn$- pairs were approximated by the
quasi-bound ``scalar deuterons". Furthermore, the use of a
``self-energy" correction to the $^3$He- propagator in ref.[24] is
not consistent with the use of a reducible $tdp$- vertex which
already includes all self-energy parts (see eq.(4)). At last,
accounting for the ISI by the modification of the $tdp$- vertices
in the pole-graphs in accordance with eq.(40) from [24] violates
T- reversal invariance, since leads to an imaginary part in the
vertices defined by the bound states wave functions only.
Generating an imaginary part in the $tpd$- vertex and violating
$T$- invariance, the authors of [24] tried to account for the
$many$-$body$ ISI effects through the $one$-$body$ (pole-type)
currents, that is not possible in principle.
\par
In [25] J.Golak et al. have shown that an (explicit) inclusion of
MEC considerably changed their previous calculations on $T_{20}$,
especially at 19.8 MeV. Their new result is shown in fig.2 by the
double-dashed curve. As it may be seen, inclusion of MEC changes
the character of their former result (dashed-dotted curve) at
small angles and comes on closer to our calculations (solid and
dashed curves in fig.2). In general, the trend of their new
calculation at $\theta < 50^0$ coincides now with our results
(compare double-dashed curve with the solid- and dashed- lines),
in a full agreement with the discussion in Sec.III (see above). As
for the higher energies, the new calculations [25] differ only
slightly from the previous ones [4], reproducing the same shape of
the curves, and we do not show them. In addition, authors of [25]
have found that after inclusion of MEC their calculations become
less sensitive to the model of NN- interaction. That is exactly
what we described in Sec.III: when one- and many-body currents are
(correctly) balanced in the total amplitude, various realistic
potentials lead to close results.
\section{ACKNOWLEDGEMENTS}
This work is supported in part by the Stichting Fundamenteel
Onderzoek der Materie (FOM) with financial support of the
Nederlandse Organisatie voor Wetenschappelijk Onderzoek(NWO). We
are  grateful to W. Gloeckle, J. Golak, H. Vitala, A. Fonseca, I.
Sick, V. Pandharipande and N. Kalantar for the stimulating
discussions and for providing numerical theoretical and/or
experimental results. In particular we would like to thank R.
Wiringa, who provided us with his calculations of the $^3$He wave
functions and various overlap integrals for the Argonne-V18
potential.
%

%
{\bf Figure captions}
\\  \\
Fig.1 \\ Gauge invariant set of  covariant diagrams for the
two-body $^3$He-photodisintegration (or radiative $dp$- capture).
\\  \\
Fig.2 \\  $T_{20}$ at $E_d = 19.6$ $MeV$. The solid (dashed) curve
presents our covariant and gauge invariant calculations (all
multipoles) with RSC (Argonne-V18) $NN$- interaction. The dotted
and dashed-dotted curves show the quantum-mechanical calculations
from ref. [2] ($E1$ multipole only) with Paris potential and ref.
[4] (all multipoles) with Bonn potential, respectively. The
double-dashed line presents new calculations from ref. [25] (the
same as in ref. [4], but with additional $MEC$- contributions).
The data are from ref. [21].
\\ \\
Fig.3 \\ Tensor analyzing power $A_{yy}$ at 29 $MeV$. The solid
curve corresponds to our present calculations with the Argonne-V18
interaction. The dashed curve reproduces the full result obtained
in ref. [4] with the Bonn potential. The experimental point is
from ref. [5].
\\  \\
Fig.4 \\ Tensor analyzing power $A_{yy}$ at 45 $MeV$. The solid
curve corresponds to our present calculations with the Argonne-V18
interaction. The dashed curve reproduces the full result obtained
in ref. [4] with the Bonn potential. The data are from ref. [4].
\\ \\
Fig.5  \\ Tensor analysing power $A_{yy}$ at 95 $MeV$. The solid
curve corresponds to our present calculations with the Argonne-V18
interaction. The dashed curve reproduces the full result obtained
in ref. [4] with the Bonn-B potential. The data are from ref.
[22].
\\ \\
Fig.6  \\ Our predictions for the tensor analysing power $A_{yy}$
at various energies. The solid curve corresponds to $E_d = 150$
$MeV$, while dashed, short-dashed, dotted and dashed-dotted
represent our results for $E_d = 200, 250, 300 and 350$ $MeV$.
\\ \\
Fig.7  \\ The cross section for the $^3$He $+ \gamma \rightarrow p
+ d$ reaction at $E_{\gamma} = 240$ $MeV$. The solid curve
corresponds to our present covariant and current conserved
calculations with Argonne-V18 interaction. The dashed
(dashed-dotted) curve shows J.-M. Laget calculations from ref.
[20] in the diagramatic approach with 2-body (3-body) MEC and FSI.
See for the data in ref. [20].
\\ \\
Fig.8  \\ The same as in fig.7, but for $E_{\gamma}=340$ $MeV$.
\\ \\
\begin{figure}
\epsfbox{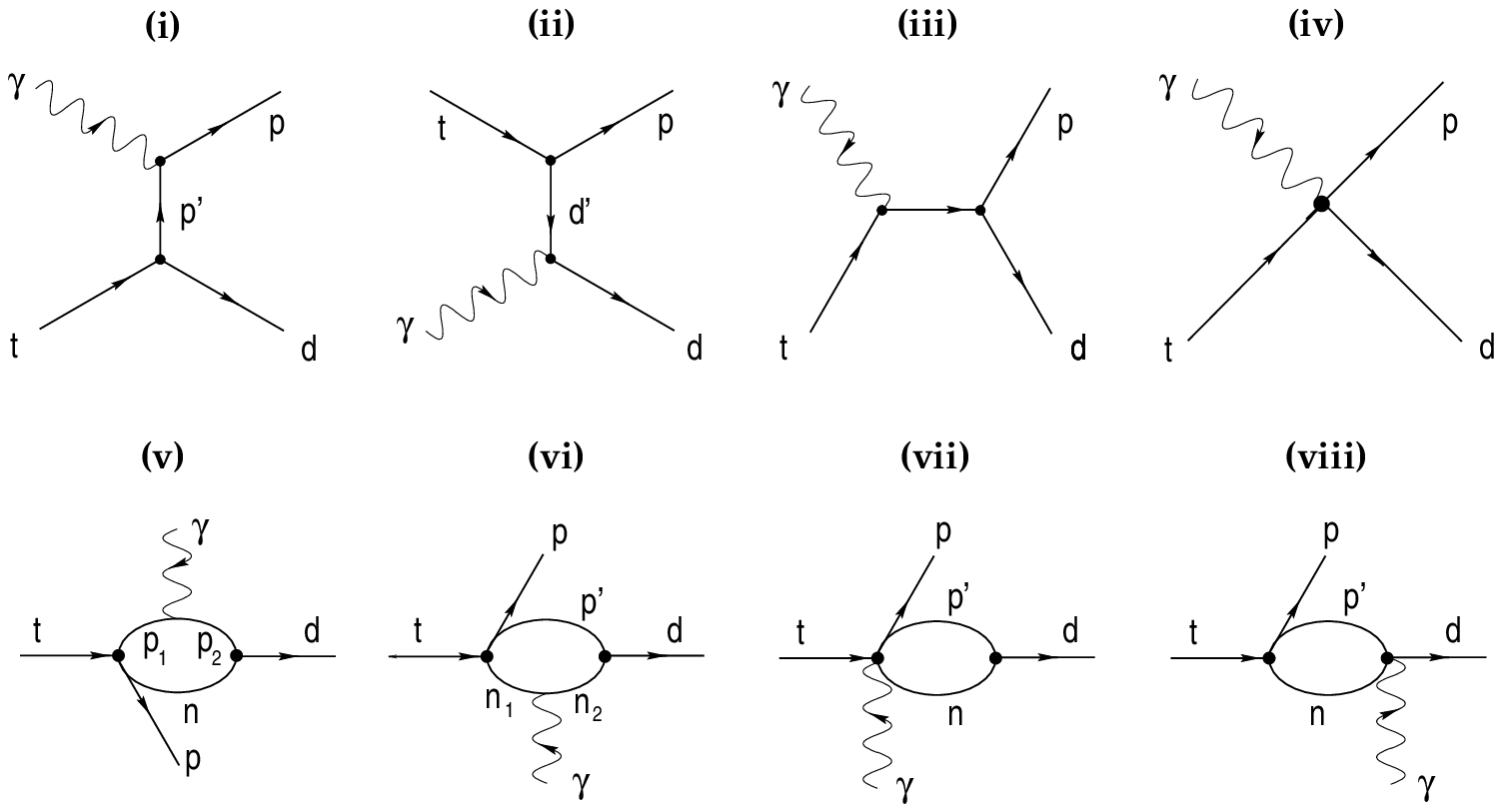} \caption{}
\end{figure}
\begin{figure}
\epsfbox{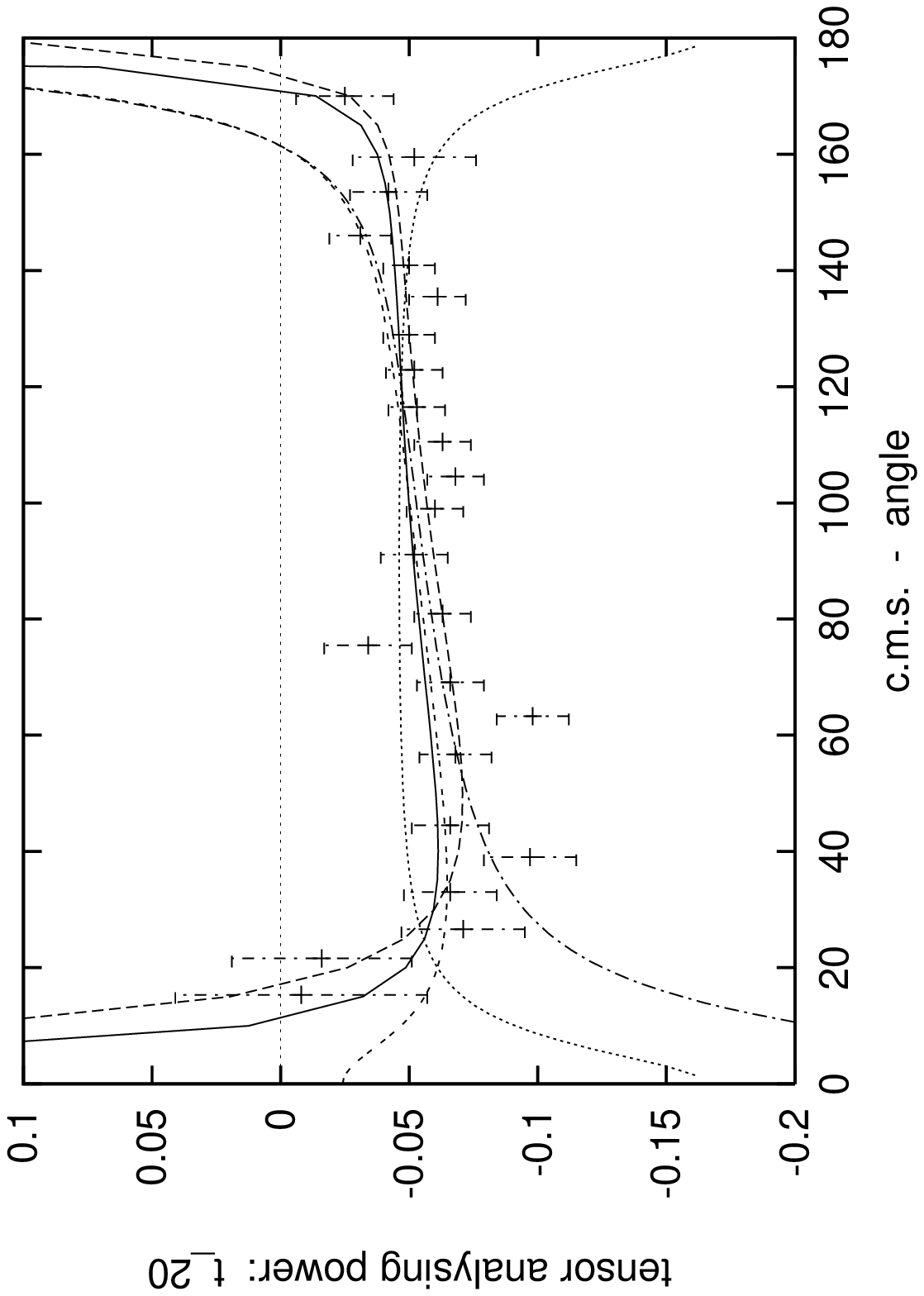} \caption{}
\end{figure}
\begin{figure}
\epsfbox{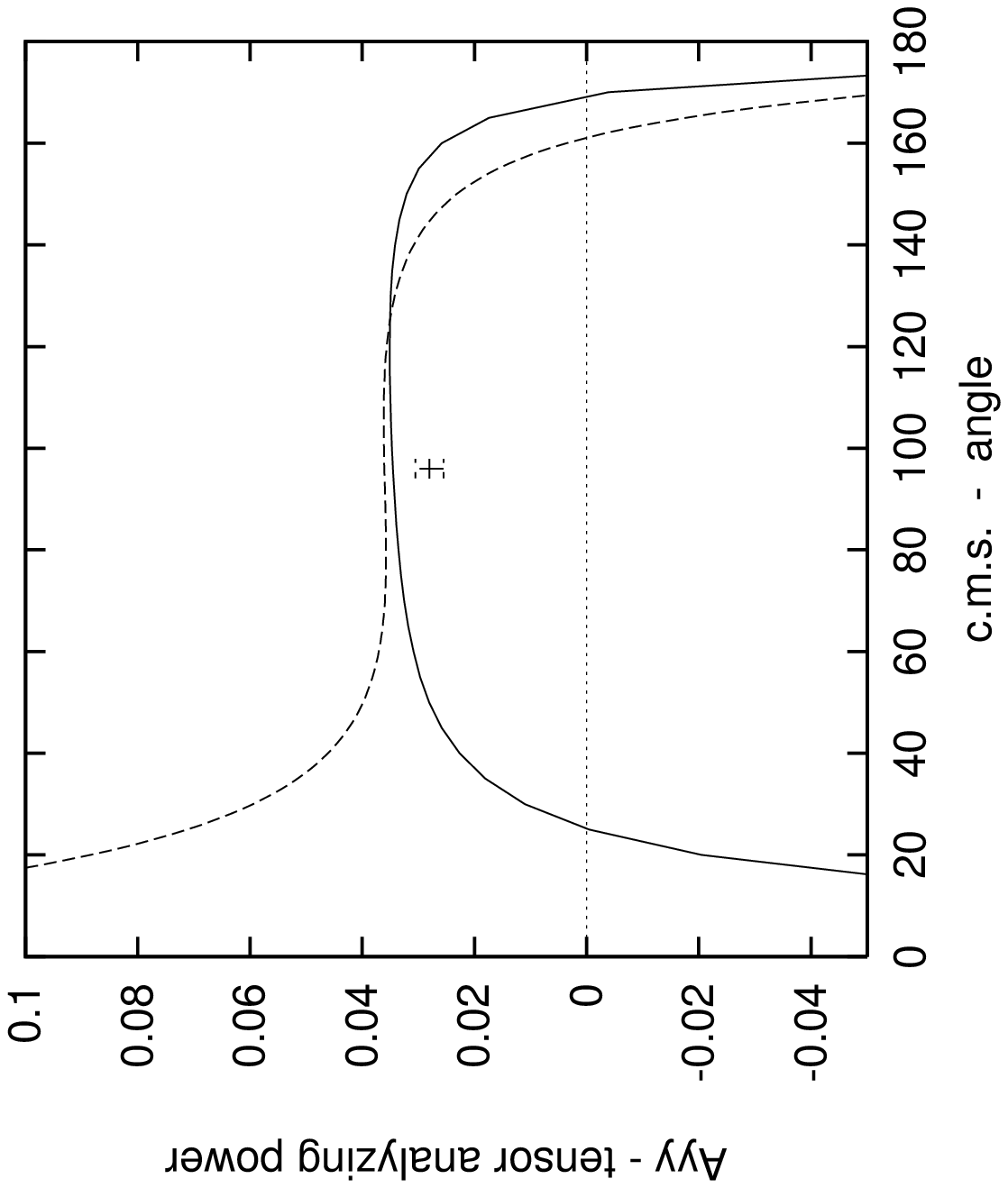} \caption{}
\end{figure}
\begin{figure}
\epsfbox{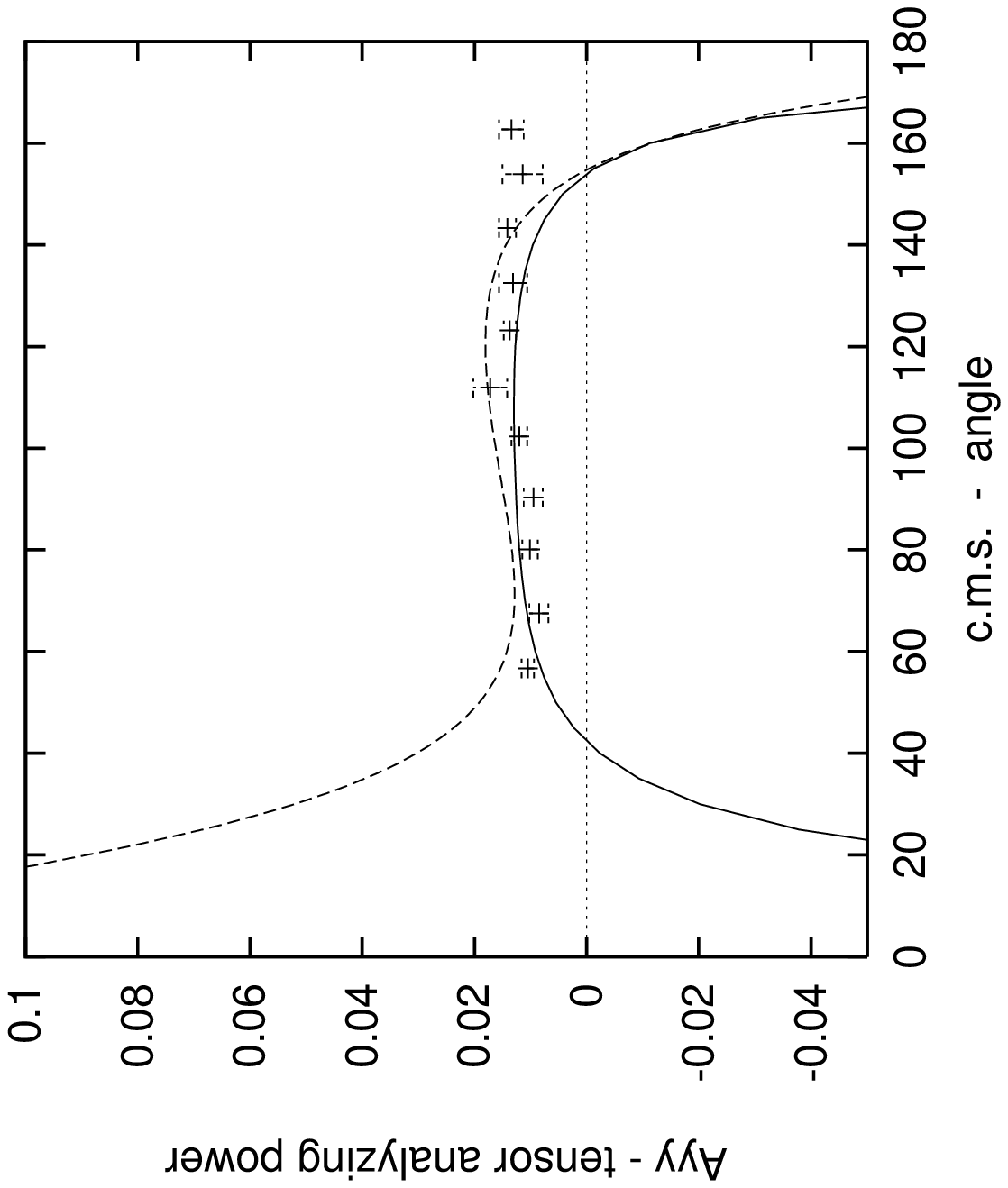} \caption{}
\end{figure}
\begin{figure}
\epsfbox{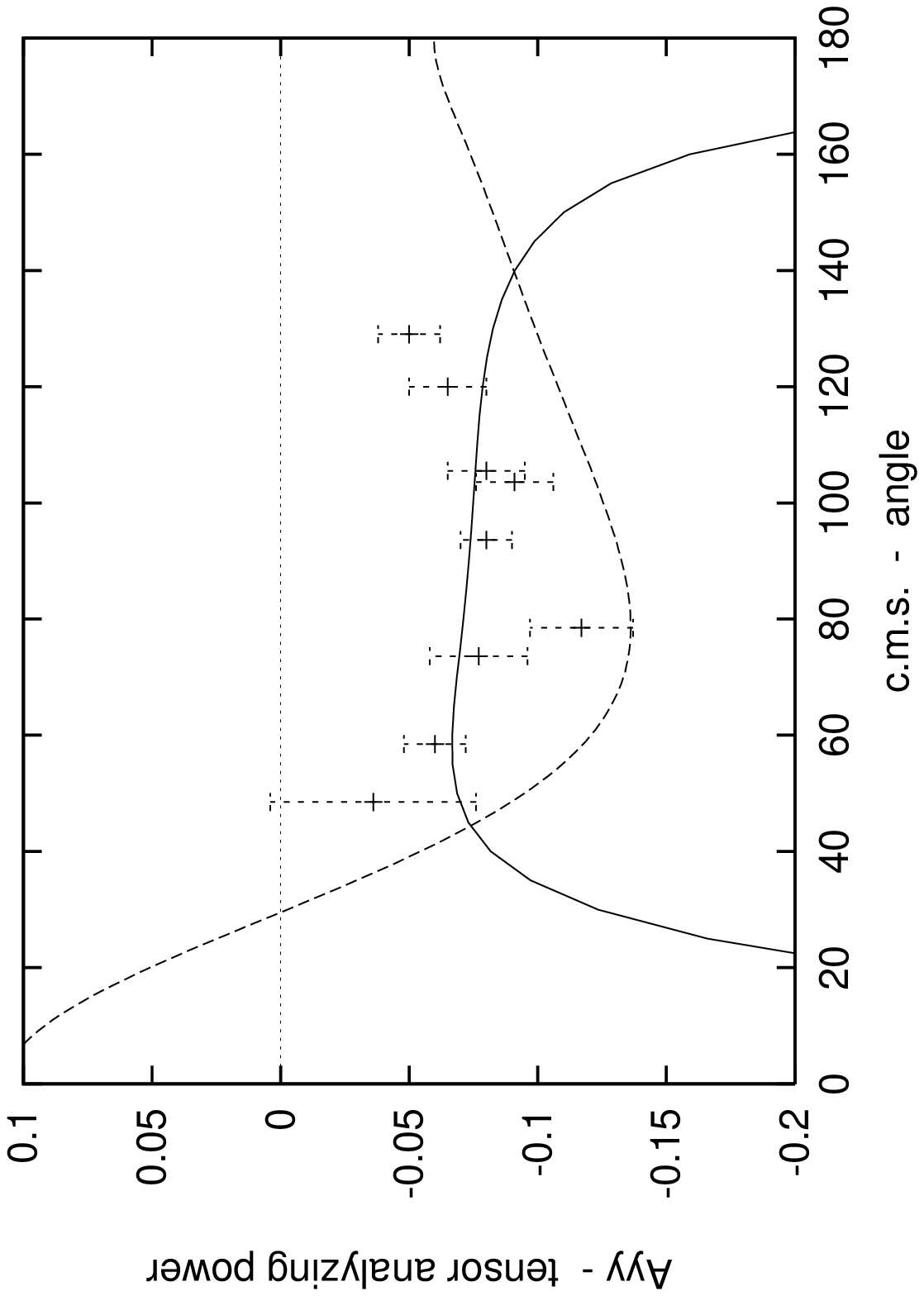} \caption{}
\end{figure}
\begin{figure}
\epsfbox{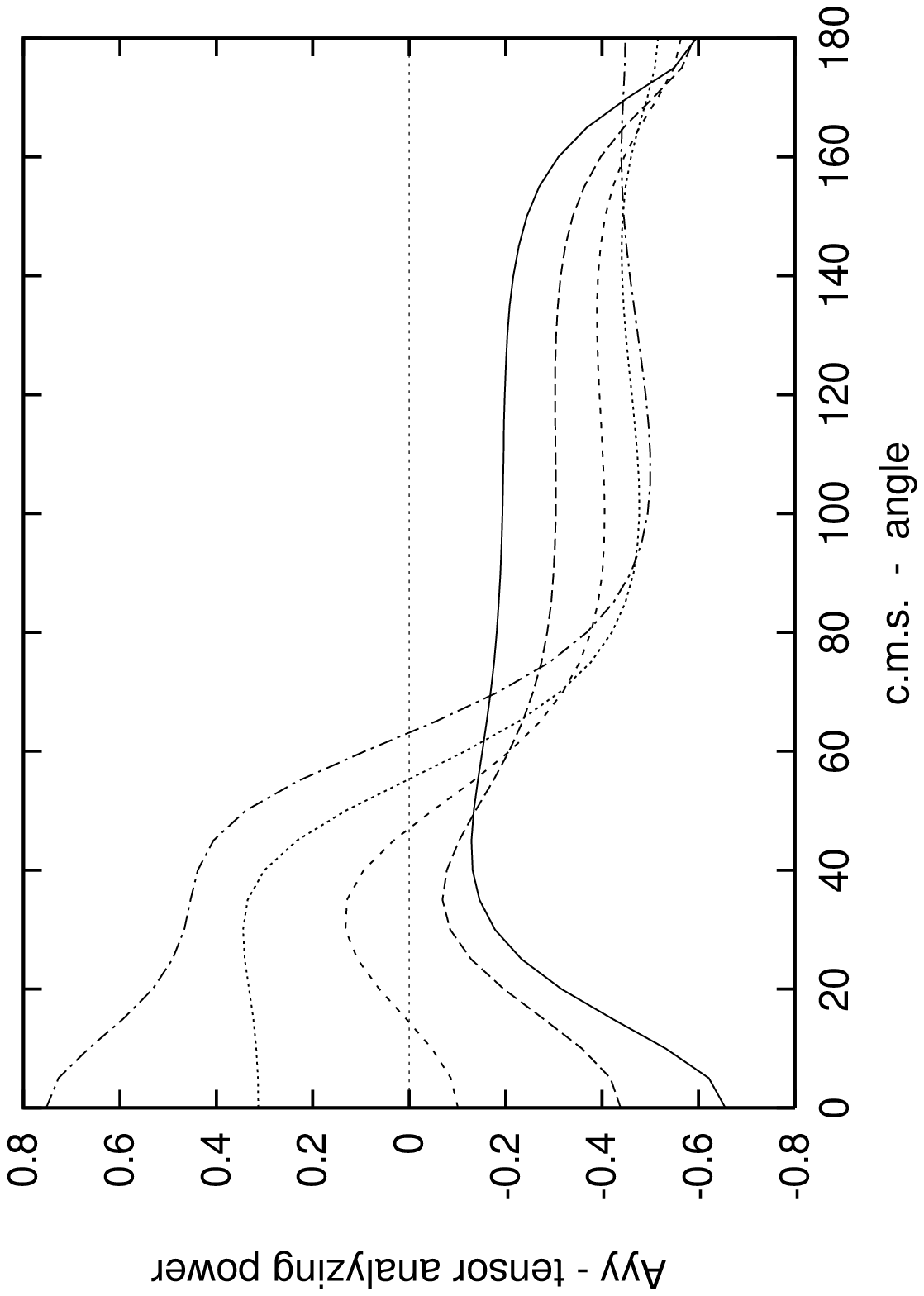} \caption{}
\end{figure}
\begin{figure}
\epsfbox{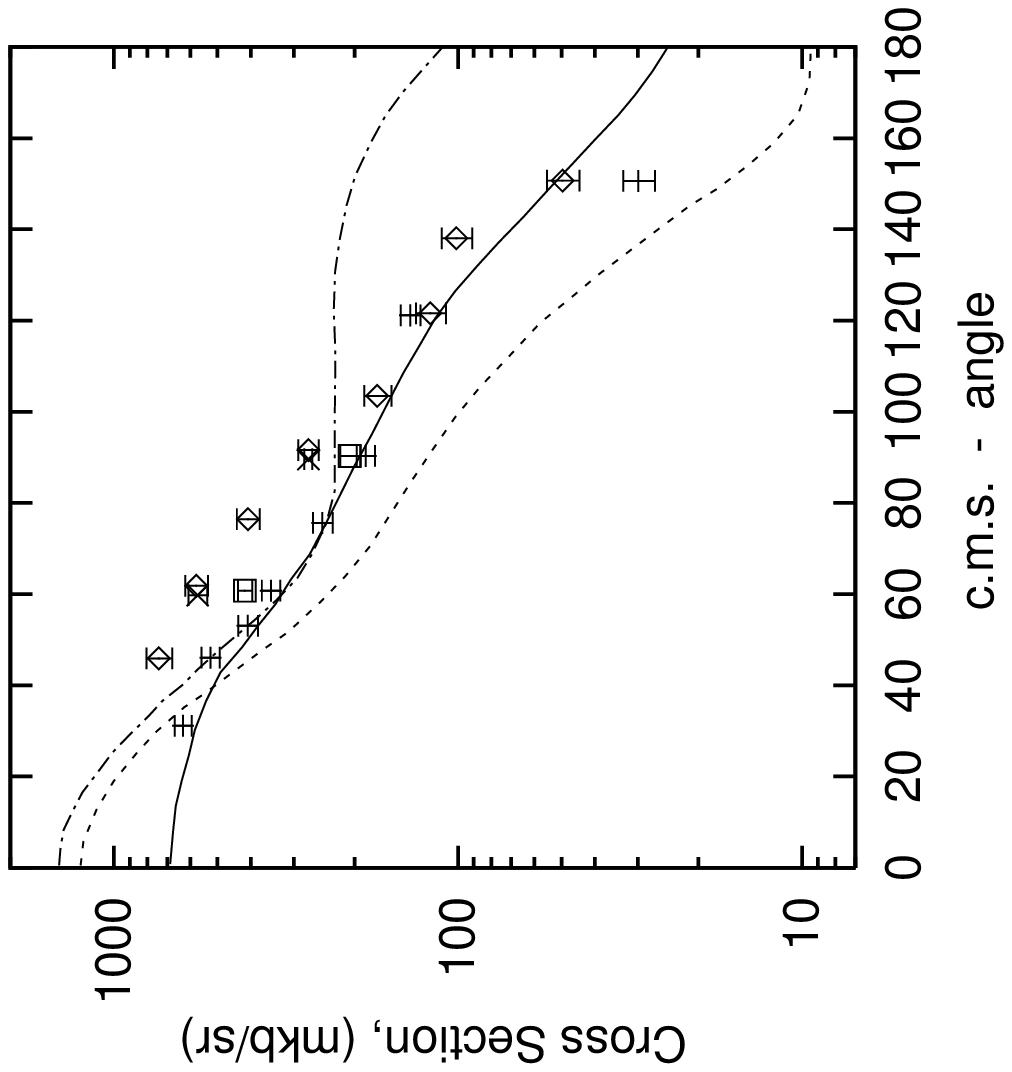} \caption{}
\end{figure}
\begin{figure}
\epsfbox{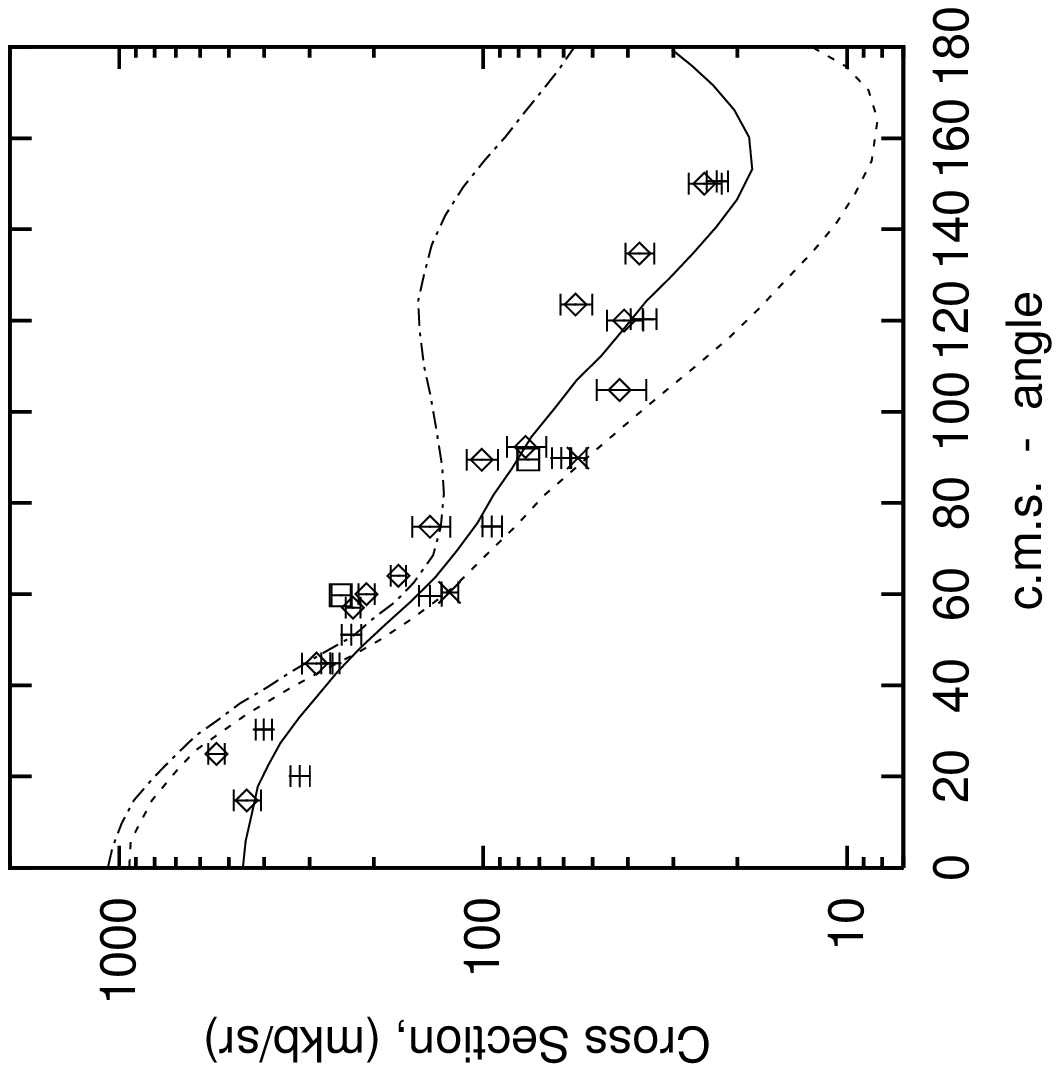} \caption{}
\end{figure}
\end{document}